# Comparative study of adatom manipulation on several fcc metal surfaces

Chandana Ghosh, Abdelkader Kara and Talat S. Rahman
*Kansas State University, Department of Physics, Manhattan, Kansas 66506*

## Abstract

For a set of fcc metals our total energy calculations, based on many body potentials, show that activation barriers for lateral manipulation of an adatom at a step edge depend on the tip/substrate composition. Of the six homogeneous systems studied, manipulation on stepped Ag(111) showed the lowest energy barrier for adatom hopping towards the tip, although the relative probability for this process was largest on Cu(111). For a representative Cu/Pt heterogeneous system we find lateral manipulation of a Pt adatom along a step on Pt(111) by a Cu(100) tip to be energetically much less favourable than the reverse case of a Cu adatom manipulated by a Pt(100) tip. In the case of vertical manipulation, atomic relaxations of the tip and its neighbouring atoms are found to be prominent and tip induced changes in the bonding of the adatom to its low coordinated surroundings help explain the relative ease with which an adatom next to a step edge or a kink site, may be pulled, as compared to that on a flat surface.

# I. Introduction

Manipulation of individual atoms, both laterally and vertically using Scanning Tunnelling Microscope (STM) has been a common affair in several research laboratories [1-9], with tempting applications including nanostructuring of surfaces and altering of chemical reactions, both of which play crucial roles in nanotechnology. Because of the vicinity of the tip to the adatom and its neighbours, questions are automatically raised about the changes in the potential energy surface and atomic relaxations brought about by the tip, and their dependencies on the structure and composition of the tip. A related question is whether heterogeneous systems can provide more facile manipulation. Additionally, there is the issue of whether defects like steps and kinks on surfaces offer better grounds for adatom manipulation. Studies of lateral manipulation on homogeneous systems [10,11] indicate that most characteristics may be explained through considerations of system energetics dictated by interatomic forces alone. Furthermore, in a related detailed study of the influence of the tip on the potential energy surface seen by Cu adatom at the step edge, not only the diffusion barriers but also the saddle points were shown to be perturbed [12]. These previous studies had focussed on lateral manipulation on homogeneous systems of Cu [11,12] and Au [10] because of their direct relevance to experimental data.

To motivate future experiments and to understand the trends in atomic manipulation for a set of metallic systems, a systematic study is needed. Such a study will help understand whether the ease of manipulation correlates with the cohesive energy of the constituents, in the case of homogeneous systems. In the present paper we have carried out a comparative study of the lateral manipulation of adatoms on six metal surfaces (Ag, Au, Cu, Pt, Ni and Pd). Additionally we have considered a heterogeneous system of Pt and Cu constituents to examine possible asymmetries in the ease of manipulation. We have also probed in detail the rationale for the differences in the vertical manipulation of adatoms from stepped, kinked and flat surfaces. Special attention has been given to atomic relaxations brought about in the presence of the tip. As shown in a recent study [13], atomic relaxations of the tip atoms and those in their vicinity may be quite pronounced.

In section II, we present a brief description of the model system. This is followed by a brief account of the theoretical techniques employed, in section III. Section IV contains results for lateral manipulation, while those for vertical manipulation are summarized in section V. Conclusions are presented in section VI.

.

## II Model Systems

The model system for the manipulation process consists of two parts: the adatom on a metal substrate, and the tip, which is a nanostructure itself. The substrate consists of 8 atomic layers in fcc stacking with (111) orientation having 10x12 atoms in each layer. Calculations performed on larger systems have the above dimensions to be sufficient to avoid problems arising from the finite size of the simulation cell. Deleting 7 and 7.5 atomic chains, respectively, from the surface layer creates a (100) microfacetted step, and kink, on the (111) surface. Lateral manipulations are carried out in this work, on the stepped surface only, while vertical manipulation is performed on flat, stepped and kinked surfaces. Sharp as well as blunt tips, having either fcc(100) or fcc(111) geometry, have been considered. Sharp tips consist of 35 metal atoms with 1 atom at the tip apex (Fig. 1a). The blunt tips have 34 atoms, leaving the tips with (100) and (111) geometry with 4, and 3 atoms, at the apex, respectively. The lateral separation between the adatom and the tip is illustrated in Fig. 1b. Since the interactions potentials for the materials considered are short ranged, the tip sizes chosen here are sufficient for examination of their effect on the potential energy surface in the vicinity of the adatom. The metals studied for lateral manipulation are Cu, Ag, Au, Pt,

Ni and Pd, so chosen because of the availability of reliable interaction potentials and their relevance to experiments.

**III Theoretical Method**

To obtain the energetics of the systems, we have carried out calculations employing interaction potentials from the embedded atom method (EAM) [14], with parameterization by Voter and Chen [15], since it is not yet computationally feasible to perform such extensive systematic studies of the potential energy surfaces from first principles electronic structure calculations. The total energy of the system consisting of the tip, the adatom and the substrate (Fig. 1a) is minimized using the conjugate gradient method [16]. The experiments that have motivated our calculations were performed at temperatures below 50K and the time scales for atomic motion are about 1ps. Thus we do not expect surface vibrations to play a significant role here and have not included dynamical contributions like those from vibrational entropy and atomic vibrational amplitudes, which might play a role at higher surface temperatures. Also, the zero point motion of atoms like Cu is not significant to change the qualitative behavior of these systems. Furthermore, Van der Waals forces are also not expected to be predominant [9], since the tip radius is only a few angstroms and the vertical separations between the tip and the adatom range from 3.0Å to about 7.5Å. During the energy minimization calculations, all atoms of the last two layers of the cell are kept rigid. Also, the four atoms at the corners of the stepped surface, as well as, those at the two corners of the lower terrace are kept rigid. Atoms in the top two layers of the tip are also kept fixed while the rest are allowed to relax to their equilibrium positions during each simulation. Beside these constraints, the results quoted in this paper are for the fully relaxed tip and surface systems.

For lateral manipulation, the total energy is calculated for each configuration of the system with the adatom placed at every point along a fine 10Å x 10Å grid consisting of 100 x 100 points, which covers part of the upper and lower terraces adjoining the step. The potential energy surface of the system was obtained from the total energy of the system for each of the $10^4$ configurations (Fig. 2). From the calculated potential energy surface, the activation energies for the adatom to diffuse in various directions were then obtained. In the case of vertical manipulation, the adatom is constrained in the Z direction (normal to the surface). This prevents the adatom from falling back to the surface when it is raised in small intervals from the surface to the tip during the energy minimization process.

In the case of vertical manipulation, the adatom is constrained in the Z direction (normal to the surface). This prevents the adatom from falling back to the surface when it is raised in small intervals from the surface to the tip during the energy minimization process. The total energy versus the Z coordinate of the adatom was obtained for several tip heights. For the case when the tip is directly above the adatom, the relaxations of the adatom, the tip atoms and the neighboring atoms are calculated using the data in the initial and final (after minimization) configurations. From these relaxations, we can determine the spacing between the tip and the adatom when repulsive forces come into play. We can also determine which of the neighboring atoms are strongly affected by the presence of the tip and consequently which atom/atoms can be pulled by the tip easily.

# IV Results

## IV.1. Lateral Manipulation

In this section we present results of a comparative study of the lateral manipulation of an adatom along a step edge on the (111) surface of six fcc metal surfaces. From examination of the changes in the activation energy barriers for adatom diffusion in the presence of the tip, we discuss the relationship of the energy barriers to the cohesive energy of the metal. We also consider the influence of the geometric relaxations of the atoms in the tip, and their nearest neighbours on the surface, on the manipulation process. Since beyond a certain separation between the tip and the substrate, the tip itself gets distorted [13], we report here results only for the lowest tip-adatom separation for which the tip remains intact. For the metals considered here 2.75 Å above the step edge appears to be a reasonable vertical height of the tip. In A recent work [12], we have shown that a sharp tip with (100) geometry is more effective as compared to the (111) tip in lowering the activation barrier. Thus, all calculations for the lateral manipulations presented here have been performed using a sharp tip of (100) geometry placed at 2.75Å above the step edge.

The fine grid method, discussed in the previous section, provides a detailed resolution of the potential energy surface [12] enabling accurate evaluation of activation energy barriers as well as the diffusion path. In the presence of the tip, the saddle point shifts with respect to its original unperturbed position and diffusion barriers are calculated with respect to this shifted saddle point [12]. These shifts in the saddle points for the six systems are listed in Table 1. The calculated activation energy barriers for the adatom on each of six metal (Ag, Au, Pt, Ni, and Pd) surfaces are also summarized in Table 1. In each case, the tip, the adatom and the substrate are composed of the same element. We also consider two heterogeneous systems in which the tip is made of one type metal (Cu or Pt) atoms while the adatom and the substrate are composed of the other. In each of the six homogeneous cases considered, the lowest diffusion barriers are found at tip-adatom lateral separations between 2.0Å - 3.0Å, which is in the range of the nearest neighbour separation. Ag is found to have the lowest barrier of 50 meV with a saddle point shift of 0.7Å, while the barrier in the direction away from the tip (op-barrier) is 215 meV. As can be seen from Table 1, however, it is Cu that shows the highest ratio of op-barrier to barrier in the presence of the tip. Note that the probability for an adatom to diffuse towards the tip depends not only on the energy barrier for the adatom to jump towards the tip, but also on the energy barrier to move away from the tip (op-barrier). The ratio of the two probabilities (to move towards and away from the tip) depends exponentially on the difference between the activation energies for these two moves. The adatom would easily follow the tip when the difference between these two energy barriers is larger than kT. Hence, the relative probability for adatom manipulation, at a step edge, by a tip is higher for Cu(111) than for Ag(111). As expected the energy barrier for the adatom to hop towards the tip is lowered and that to diffuse in the opposite direction is raised in the presence of the tip. The effect is, however, not uniform.

In particular, we find an anomaly in the case of Pt, for which the energy barrier for the adatom to diffuse in the direction of the tip is found to be higher than that in the absence of the tip. Looking into the relaxation patterns of the adatom and its neighbours for this case, we find a significantly large relaxation of the tip apex atom towards the adatom and the adatom towards the tip, as will be discussed in details in the next section. This anomaly could be due to the interaction potentials from EAM, which are not as accurate for Pt as they are for the other metals. But nevertheless, it is an interesting result and needs deeper investigation. If it is in fact due to the EAM potentials then one should note that similar calculations on Pt metal provide a good check of the reliability of these potentials for this metal.

To find a trend in the behaviour of these metals with respect to adatom manipulation we have compared the energetics of the system in the presence of the tip to the cohesive energy. We have thus summarized in Table 2 the ratio of the barrier, op-barriers, and their difference to the cohesive energy. We have also included in the table, the relative enhancement of the op-barriers to the barrier. There is a definite trend in the variation of the energy barrier for the adatom to move in the direction of the tip with the cohesive energy of the metal, indicating that atomic manipulation is easier on Ag and Cu than the others. In fact, Cu has a small edge over Ag. Interestingly, not much can be concluded from the variation of the op-barrier with the cohesive energy. Of course, these conclusions are derived within the context and accuracy of EAM potentials. Nevertheless, the trend in the six metals is interesting.

In the case of the two heterogeneous systems that we have considered, we find that a Pt tip lowers the barrier for a Cu adatom to hop towards it on a Cu substrate to 180 meV which is 3 times larger than the barrier found for the homogeneous Cu system (see Table I). The Pt tip is thus not as effective as a Cu one for manipulation purposes. However, in the reverse case of a Cu tip on a Pt substrate with a Pt adatom, the barrier is found to be slightly lower than that in a full Pt system. To understand this very interesting result of the asymmetric behaviour in the Pt/Cu and Cu/Pt systems, we turn to considerations of the relaxations of the adatom, its nearest neighbour atoms and the tip atoms.

**IV.2. Relaxations of adatom, neighbouring atoms and tip atoms - Cu/Pt and Pt/Cu systems**

The relaxations of the adatom, its neighbouring atoms and the tip atoms are discussed below. To make a comparison between the homogeneous and heterogeneous systems, we have included the cases for the pure Cu/Cu and Pt/Pt systems. In Fig. 3 we show the labelling of the adatom, its nearest neighbours (1-11) and the tip atoms (T1-T5). From Fig. 4 it can be seen that at the tip height of 2.75Å (for the pure Cu and the pure Pt systems), 3.0Å for the Cu tip on Pt system, and 3.5Å for the Pt tip on Cu system. The lateral separation between the tip and the adatom has been taken at the value when the barrier is at the minimum (see Table 1). The tip apex atom T1, as well as most of the tip atoms, shows a downward relaxation towards the adatom (negative value). The adatom relaxes upwards towards the tip and the nearest neighbours of the adatom also relax upwards providing coordination to the adatom, and lowering the total energy of the system. As Fig. 4 indicates, the downward relaxation of the tip atom is largest for the Cu tip acting on the Pt surface. Since there is a collective effect of the nearest neighbours on the relaxation of the adatom, it is interesting to examine the sum of the nearest neighbour distance between the adatom and its neighbours for the configuration in which the energy is lowered in the vicinity of the tip, as well as for the configuration in which the adatom is far away from the tip (Table 3).

From the above data we find that the total NN distance is in fact smaller (atoms are closer to each other) when the tip is farthest as compared to when it is in the vicinity of the adatom. We can thus conclude that even though the nearest neighbours of the adatom relax towards the tip, they do not provide as much coordination as they do when the tip is far away from the adatom. When the tip gets close to the adatom, the attraction of the adatom towards the tip wins over those for its neighbours on the surface making the corresponding bonds with the surface neighbours looser. The competing effect of the tip atoms and the surface nearest neighbours on the adatom means that we have to look deeper to explain the reason for the lowering of the activation energies in the presence of the tip than simple coordination effects. Next we look at the individual binding energies of the adatom, its nearest neighbours and the tip atoms (Table 4).

From table 4 it can be seen that the binding energies of the surface atoms are lower when in the vicinity of tip (when the adatom is ready to hop), as compared to the binding energies when away from the tip. This indicates that the adatom, its nearest neighbours and the tip atoms (especially the apex) strongly interact with each other, causing the lowering of the energy at this position (in the vicinity of the tip). For the pure Cu system, the binding energies of the adatom and the tip apex when close to and away from the tip are not so dramatic. When the Cu tip is replaced with a Pt tip, the tip apex does not interact strongly with the Cu system. Thus there is no significant effect created by its presence. But for the pure Pt system, there is a significant difference in the binding energies of the adatom and the tip apex, showing a very strong interaction to bind with each other. When the tip is replaced with a Cu tip, only the tip apex is seen interacting with the Pt system, but the Pt adatom and its closest neighbours, atom #2 and #3, do not interact with the tip.

**IV.3. Vertical Manipulation**

As discussed in an earlier publication [12], in vertical manipulation of an adatom on flat, stepped and kinked Cu(111) surfaces, a Cu(100) blunt tip is found to be more effective in the manipulation process, as compared to a sharp tip because of the presence of a larger number of atoms at the tip apex which produces larger relaxations on the adatom as well as the atoms in its vicinity. The results on the stepped Cu(111) surface show that at a certain height (tip-substrate distance), the barrier to jump to the tip vanishes and the adatom can be pulled to the tip apex [5,6,12]. Calculations performed on flat, stepped and kinked surfaces clearly show that as the tip is lowered the energy profile of the system changes from a double-well like curve, to one in which the barrier hump disappears and the two wells merge into a single one [5,12]. In support of our earlier statement that relaxations of neighboring atoms play a major role in the above results for the three surfaces, we make a comparison of the sum of the displacements and that of the individual energies of the nearest neighbor atoms as the adatom is raised from the surface to the tip apex. Fig. 5, which shows the comparison of the sum of displacements of the nearest neighbor atoms for the three surfaces, is for the case of a tip height of 5.3Å, at which a zero barrier is experienced by the kinked surface. This figure clearly indicates that for the kinked surface the perturbations caused by the tip to the neighboring atoms is higher than that for the stepped and the flat surface, making it easier for an adatom pick up. Similarly, we find the stepped surface to be more perturbed by the tip as compared to the flat surface.

Table 5 gives a direct comparison of the barriers for the three types of surfaces, the flat, stepped and kinked surfaces at different heights. It is seen that the barrier for the kinked surface reaches zero at a tip height of 5.3Å, whereas the tip has to be brought a little closer to pick the adatom from a stepped surface (5.2Å) and still closer for a flat surface (4.92Å). To illustrate these effects further, we show in Fig. 6 the sum of binding energies of the nearest neighbor atoms, as the adatom is raised from the surface to the tip apex. Thus, this energy for the case of the kinked surface is greater than the other two, the flat surface being the lowest. This illustrates that the perturbation in the local environment of the low coordinated kinked and stepped surfaces is more pronounced. A recent theoretical work [13] also concluded that at very small tip-adatom separations, during contact formation, several effects acting simultaneously finally lead to the pick up of the adatom.

**IV.4. Atomic relaxations in the case of vertical manipulation on flat, stepped and kinked surfaces**

To understand and study the process of vertical manipulation in more detail, we have studied the relaxations of the adatom as well as its neighbouring atoms in the presence of sharp and blunt tips with (100) and (111) geometry. As will be seen, the combined relaxations of the neighbouring atoms provide a rationale for understanding the results for vertical manipulation on flat, stepped and kinked surfaces. We find that, at a tip-adatom separation of 2.5Å, the adatom and atoms labelled 2 and 3 (Fig. 3) relax towards the tip by more than 0.1Å. Atoms 7, 8 and 10, on the other hand, relax much less because they are further away from the tip. The tip apex atom, together with the other atoms in the tip, relaxes upward into the tip, seeking higher coordination other tip atoms. At higher tip-adatom separations, the relaxations of the adatom and its neighbours are negligible and those of the tip atoms are in an upward direction towards the tip structure itself, forming a more compact tip, or a tightening effect.

If we look at the corresponding relaxations that occur in the presence of a (100) blunt tip, we find that again the adatom and its neighbours, atoms #2, 3 have a stronger attraction (about 0.65Å) towards the tip at a tip height of 3.0Å, which is much higher than that seen for a sharp (100) tip. Most of the tip atoms relax downward towards the adatom and its neighbours at larger height of the tip, 3.0Å and 3.5Å as compared to 2.75Å for the sharp tip, suggesting that the blunt tip is more efficient for vertical manipulation than the sharp tip.

Moving on to the case of the sharp and blunt (111) tips, we find that in the presence of a (111) sharp tip, at a tip-adatom separation of 2.5Å are somewhat different from those for the (100) tip, as relaxations are larger (up to 0.2Å). Here too the adatom as well as all neighbouring atoms relax towards the tip. Again, atoms 2 and 3 show similar trends, and atoms 7, 8 and 10 are affected similarly. The tip apex atom relaxes downward towards the adatom, as if the tip has a loosening effect. This was not the case for the (100) sharp tip and this could be the reason for the larger upward relaxation of the adatom and its neighbours at this height. At larger tip-adatom separations, the relaxations of the adatom and its neighbours are negligible.

As for the blunt (111) tip, we find that for a low tip height of 3.0Å, the adatom and its neighbours show much larger relaxations towards the tip as compared to a sharp (111) tip. At this height, the 3 apex atoms of the blunt (111) tip relax downward, towards the adatom, which eventually leads to the pick up of the adatom. These relaxations of the tip apex atoms, the adatom and its neighbouring atoms give us a clearer picture of which atoms are affected by the tip presence and thus assist in the manipulation process by providing coordination effects.

## VI. Conclusions

In summary, we have examined the trends in the lowering of the activation barrier for the adatom to hop towards the tip during lateral manipulation for six metals and two heterogeneous cases of a Cu(100) tip on a Pt(111) stepped surface with an adatom, and a Pt(100) tip above a Cu(111) stepped surface with a Cu adatom. The lowest barrier is reached at a lateral separation between 2.0Å -3.0Å. The role of cohesive energy, which is specific of each metal and defined as the energy to extract an atom out of its crystal lattice, is necessary to be considered, along with the energy barriers of the atom to hop from one site to the other, to make a direct comparison between the different metals studied. Thus, the ratios of the barriers and op-barriers to the cohesive energy help in explaining this interesting result of the changes in the potential energy surface due to the interatomic potentials. Among all the transition metals studied with the same tip-substrate elemental composition, Ag is found to have the lowest barrier of 50 meV with a saddle point shift of 0.7Å. However, even though Cu had the next lowest barrier of 61.7 meV with a saddle point shift of 0.6Å, it is Cu that is the easiest to manipulate. This is concluded from the ratios of the op-barrier to the barrier in the presence of the tip and also the ratios of difference in the op-barrier and barrier to the cohesive energy and to the barrier in the absence of the tip. These ratios indicate that the probability of the adatom to be pulled towards the tip is the highest in the case of Cu as compared to the other metals. We find an anomaly in the case of the Pt metal where the presence of the tip is found to hinder the motion of the adatom along the step edge as compared to when in the absence of the tip. This could be due to the EAM potentials that are not as accurate for the Pt metal. However, we cannot rule out the fact that it is an interesting result and needs deeper investigation. For the heterogeneous cases, of a Pt tip on a Cu surface with a Cu adatom and a Cu tip on a Pt surface with a Pt adatom, the lowering of the barrier is found to be due to the interaction between the tip and the adatom. This is indicated by the binding energies of the adatom, its nearest neighbors and the tip atoms. The Cu tip is more effective in lowering the barrier on a Pt surface with a Pt adatom as compared to a Pt tip. However, for the reverse case, a Pt tip is not as effective in lowering the barrier for a Cu surface with a Cu adatom as compared to a Cu tip. The shift in the saddle point is insignificant and is found to be 0.3Å for the Pt tip on Cu surface with a Cu adatom and 0.1Å for a Cu tip on a Pt surface with a Pt adatom.

During vertical manipulation, in the absence of tip the energy required to pull an adatom from a flat surface (2.47eV) is lower than that from a stepped (3.21eV) and a kinked surface (3.53eV). As the tip is lowered to facilitate manipulation, the changes in the surface energetics are such that it is easier to extract an adatom from a stepped or a kinked surface than from a flat one, implying that the tip affects the low coordinated local environment of the adatom, which assists in its extraction. In addition, for specific tip heights, there is a floating region on the kinked and stepped surfaces in which the adatom is found to be equally attracted to the substrate and to the tip apex atoms.

The relaxations of the tip apex atoms, the adatom and the neighboring atoms show that a blunt tip is more effective in vertical manipulation. The comparisons of the sum of displacements of the nearest neighbor atoms as the adatom is raised from the surface to the tip apex as well as the comparison of the sum of the individual energies of the nearest neighbor atoms as the adatom is raised from the surface to the tip apex indicate the contribution of the local environment to the lowering of the barrier for vertical manipulation by providing coordination effects. From this detailed study, we obtain useful and interesting information about adatom manipulations for homogeneous and heterogeneous systems, the modifications to the potential energy surface in the presence of the tip and relaxations of the atoms assisting in the process, thus providing us with important factors that can help in the manipulation process.

**Acknowledgements**

This work was partially supported by the US National Science Foundation Grant number EEC-0085604.

**Table 1** Comparison of energy barriers for lateral manipulation on stepped fcc (111) metals for a tip height 2.75Å above the step edge. Also reported is the bulk nearest neighbour distance (bulk nnd).

| Metal | Barrier in the absence of tip (meV) | Barrier in the presence of tip (meV) | Op-barrier (meV) | Shift in saddle point (Å) | Lateral separation between tip and adatom (Å) when barrier is lowest (bulk nnd) |
|---|---|---|---|---|---|
| Cu | 267 | 61.7 | 512 | 0.6 | 2.55 (2.55) |
| Ag | 215.5 | 50.1 | 283.6 | 0.7 | 2.6 (2.89) |
| Ni | 308.3 | 131.3 | 386.4 | 0.45 | 2.4 (2.49) |
| Pd | 355 | 190.1 | 426.9 | 0.4 | 2.5 (2.75) |
| Au | 416.7 | 324.3 | 414.9 | 0.12 | 2.6 (2.88) |
| Pt/Cu | 267 | 180 | 300 | 0.3 | 2.6 |
| Cu/Pt | 620.8 | 590 | 670 | 0.1 | 2.1 |

**Table 2** Comparison of energy barriers and cohesive energy

| Metal | Cohesive Energy (meV) | Ratio of Barrier in presence of tip to Cohesive Energy | Ratio of Op-barrier in presence of tip to Cohesive Energy |
|---|---|---|---|
| Ag | 2960 | 0.0169 | 0.0958 |
| Cu | 3500 | 0.0176 | 0.1047 |
| Ni | 4435 | 0.0296 | 0.0871 |
| Pd | 3936 | 0.0483 | 0.1084 |
| Au | 3780 | 0.0858 | 0.1097 |

**Table 3** Total sum of the distance between adatom and its nearest neighbours

| System | Total NN distance when tip is closest | Total NN distance when tip is farthest |
|---|---|---|
| Pure Cu | 10.275 | 9.876 |
| Pt tip on Cu | 9.916 | 9.885 |
| Pure Pt | 10.898 | 10.451 |
| Cu tip on Pt | 10.724 | 10.442 |

**Table 4** Binding energies of the adatom, its nearest neighbours (when in the vicinity of the tip) and the tip atoms.

| System | Atom number | Binding energy in the vicinity of the tip (eV) | Binding energy at position away from the tip (eV) |
|---|---|---|---|
| Pure Cu system | **Adatom** | **-2.685** | **-2.525** |
| | #2 | -2.980 | -2.855 |
| | #3 | -3.129 | -2.988 |
| | #7 | -3.386 | -3.299 |
| | #8 | -3.402 | -3.307 |
| | #10 | -3.232 | -3.127 |
| | **T1 (Tip apex)** | **-2.756** | **-2.563** |
| | T2 | -2.703 | -2.690 |
| | T3 | -2.630 | -2.646 |
| | T4 | -2.691 | -2.646 |
| | T5 | -2.610 | -2.595 |
| Pt tip on Cu system | **Adatom** | **-2.615** | **-2.524** |
| | #2 | -3.023 | -2.895 |
| | #3 | -4.349 | -4.224 |
| | #7 | -3.378 | -3.299 |
| | #8 | -3.401 | -3.312 |
| | #10 | -3.234 | -3.128 |
| | **T1 (Tip apex)** | **-1.845** | **-1.853** |
| | T2 | -4.607 | -4.605 |
| | T3 | -4.499 | -4.498 |
| | T4 | -4.491 | -4.502 |
| | T5 | -4.210 | -4.207 |
| Pure Pt system | **Adatom** | **-4.833** | **-4.291** |
| | #2 | -5.067 | -4.851 |
| | #3 | -5.405 | -5.106 |
| | #7 | -5.591 | -5.468 |
| | #8 | -5.646 | -5.491 |
| | #10 | -5.388 | -5.254 |
| | **T1 (Tip apex)** | **-4.961** | **-4.425** |
| | T2 | -4.839 | -4.695 |
| | T3 | -4.644 | -4.612 |
| | T4 | -4.778 | -4.567 |
| | T5 | -4.529 | -4.499 |
| Cu tip on Pt system | **Adatom** | **-3.894** | **-4.292** |
| | **#2** | **-4.209** | **-4.837** |
| | **#3** | **-4.175** | **-4.182** |
| | #7 | -5.559 | -5.470 |
| | #8 | -5.560 | -5.464 |
| | #10 | -5.367 | -5.252 |
| | **T1 (Tip apex)** | **-7.274** | **-6.334** |
| | T2 | -2.508 | -2.568 |
| | T3 | -2.539 | -2.554 |
| | T4 | -2.601 | -2.538 |
| | T5 | -2.442 | -2.443 |

**Table 5** Comparison of energy barriers for vertical manipulation on flat, stepped and kinked Cu(111) surfaces for different tip heights.

| Surface<br>Tip<br>Height (Å) | Flat<br>Barrier (eV) | Stepped<br>Barrier (eV) | Kinked<br>Barrier (eV) |
|---|---|---|---|
| 12.5 | 2.47 | 3.21 | 3.53 |
| 8.0 | 2.26 | 2.13 | 2.42 |
| 7.5 | 2.09 | 1.45 | 1.83 |
| 6.5 | 1.51 | 1.00 | 1.09 |
| 5.5 | 0.53 | 0.29 | 0.24 |
| 5.3 | 0.34 | 0.06 | 0.00 |
| 5.2 | ___ | 0.00 | ___ |
| 5.1 | 0.19 | ___ | ___ |
| 4.92 | 0.00 | ___ | ___ |

## Figure Captions

Figure 1a: An example of an adatom against a step in the presence of a (100) tip. The square represents the grid on which the total energy of the system is calculated as explained in the text.
Figure 1b: Lateral separation between tip and adatom

Figure 2: Top view of the grid along which the adatom is placed in steps of 0.1Å. Dark circles are atoms of the top terrace of the stepped surface and the lighter circles are the atoms of the layer beneath.

Figure 3: Labels for adatom (1) and its neighbouring atoms. Below Left – Labels for the (100) tip atoms. Below Right – Labels for the (111) tip atoms. The apex atom in each case is labelled T1.

Figure 4: Relaxations of the adatom, its neighbouring atoms and the tip atoms for the four systems, Pt tip on Pt surface, Cu tip on Pt surface, Cu tip on Cu surface and Pt tip on Cu surface, at a tip height of 2.75Å, and lateral separation at the value when the barrier is minimum (from Table 1).

Figure 5: Comparison of the sum of the displacements of the nearest neighbour atoms as the adatom is raised from surface to the tip apex.

Figure 6: Comparison of the sum of the individual energies of the nearest neighbour atoms as the adatom is raised from the surface to the tip apex.

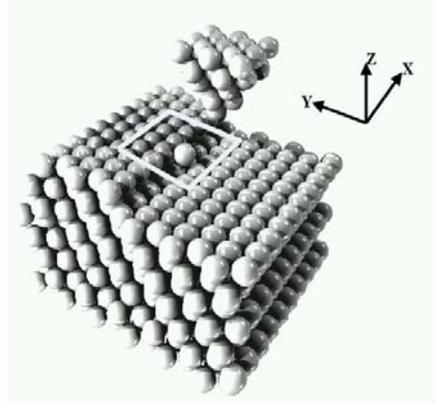

**Fig. 1a** An example of an adatom against a step in the presence of a (100) tip. The square represents the grid on which the total energy of the system is calculated as explained in the text.

.

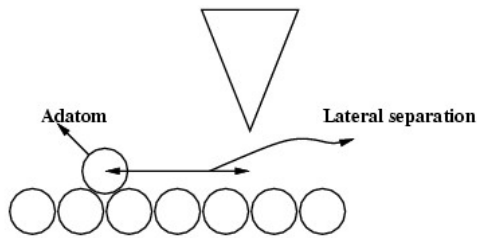

**Fig. 1b** Lateral separation between tip and adatom

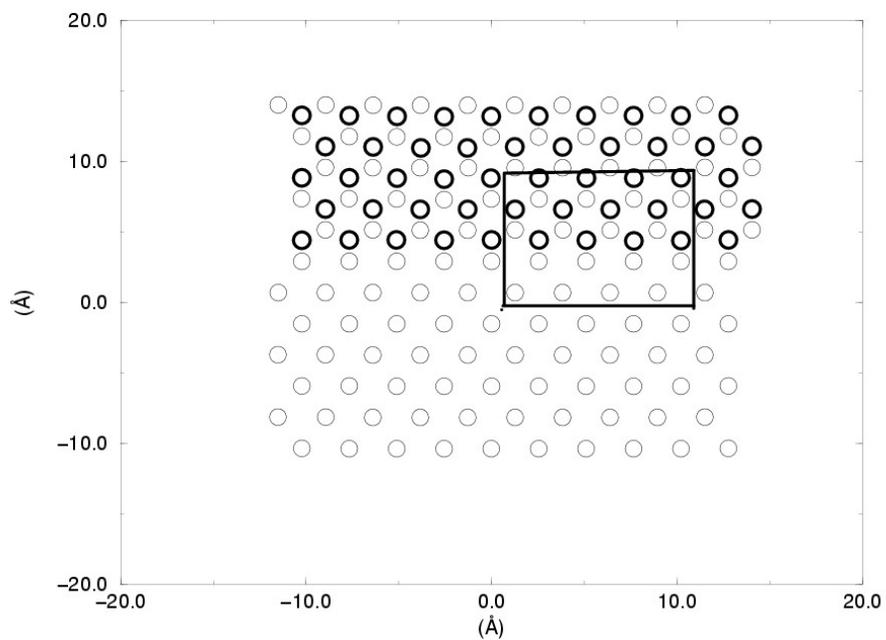

**Fig. 2** Top view of the grid along which the adatom is placed in steps of 0.1Å. Dark circles are atoms of the top terrace of the stepped surface and the lighter circles are the atoms of the layer beneath.

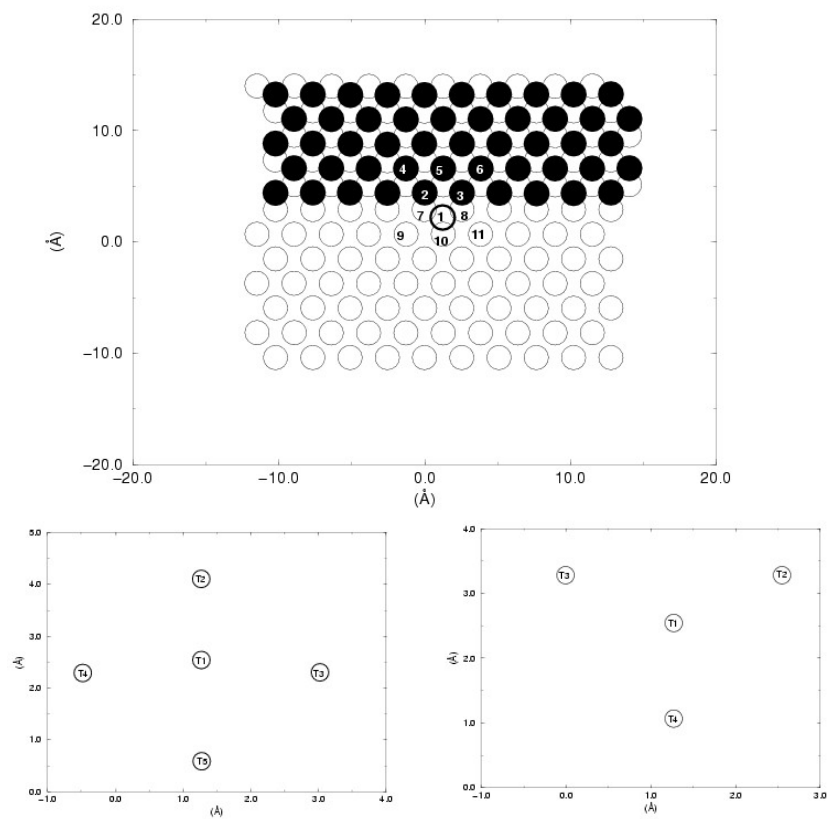

**Fig. 3** Labels for adatom (1) and its neighboring atoms. Below Left – Labels for the (100) tip atoms. Below Right – Labels for the (111) tip atoms. The apex atom in each case is labelled T1.

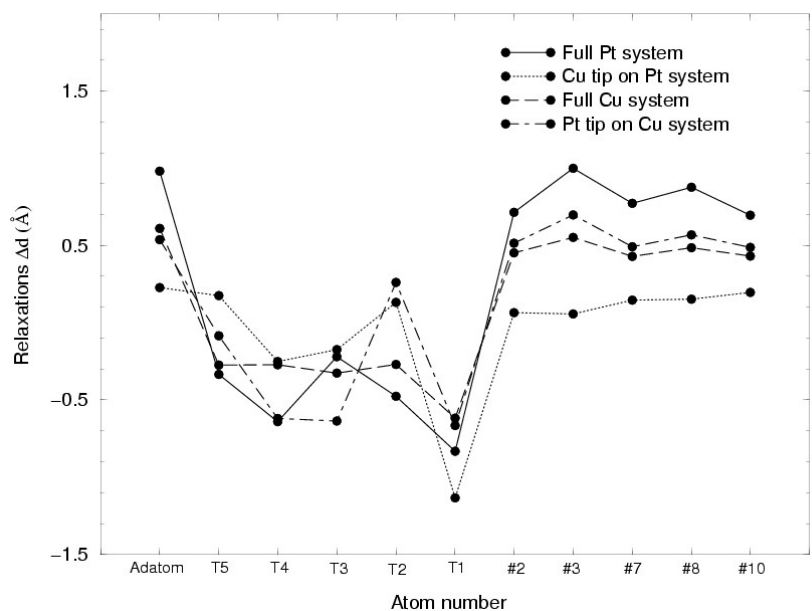

**Fig. 4** Relaxations of the adatom, its neighbouring atoms and the tip atoms for the four systems, Pt tip on Pt surface, Cu tip on Pt surface, Cu tip on Cu surface and Pt tip on Cu surface, at a tip height of 2.75Å, and lateral separation at the value when the barrier is minimum, from Table 1.

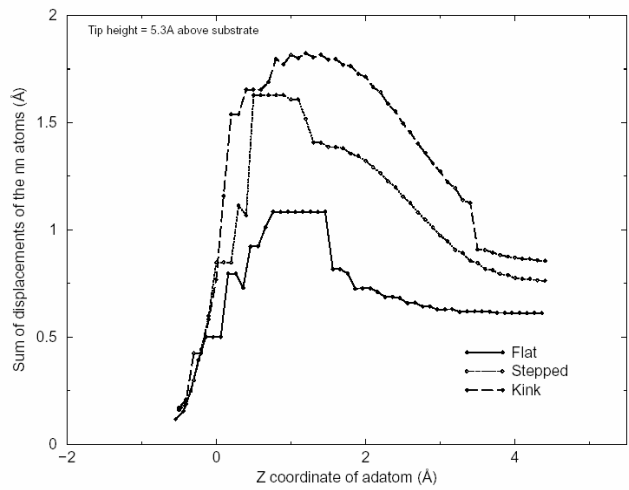

**Fig. 5** Comparison of the sum of the displacements of the nearest neighbour atoms as the adatom is raised from surface to the tip apex.

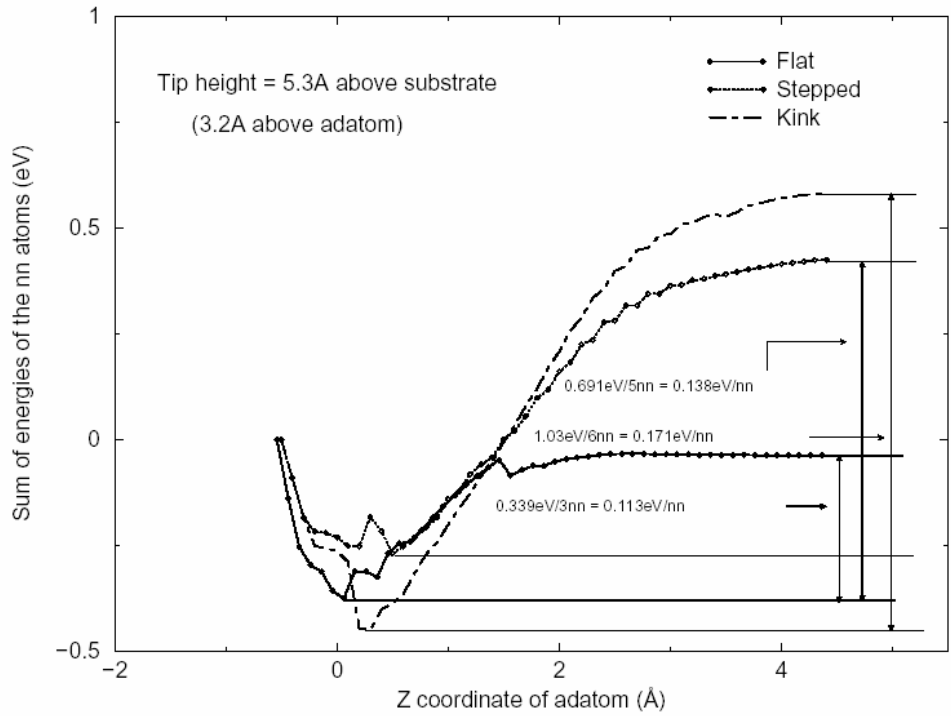

**Fig. 6** Comparison of the sum of the individual energies of the nearest neighbour atoms as the adatom is raised from the surface to the tip apex.